\documentclass[sn-mathphys,Numbered]{sn-jnl}

\usepackage{graphicx}%
\usepackage{multirow}%
\usepackage{amsmath,amssymb,amsfonts}%
\usepackage{amsthm}%
\usepackage{mathrsfs}%
\usepackage[title]{appendix}%
\usepackage{xcolor}%
\usepackage{textcomp}%
\usepackage{manyfoot}%
\usepackage{booktabs}%
\usepackage{algorithm}%
\usepackage{algorithmicx}%
\usepackage{algpseudocode}%
\usepackage{listings}%
\usepackage{adjustbox}

\renewcommand{\figurename}{}
\renewcommand{\tablename}{}

\theoremstyle{thmstyleone}%
\theoremstyle{thmstyletwo}%

\theoremstyle{thmstylethree}%

\begin{document}

\title[Article Title]{Subgiants in NGC 188 Reveal that Rotationally Induced Mixing Creates the Main Sequence Li-Dip}

\author*[1,2]{\fnm{Qinghui} \sur{Sun}}\email{qinghuisun1@gmail.com}

\author[3]{\fnm{Constantine P.} \sur{Deliyannis}}\email{cdeliyan@indiana.edu}

\author[4]{\fnm{Barbara J.} \sur{Anthony-Twarog}}\email{bjat@ku.edu}

\author[4]{\fnm{Bruce A.} \sur{Twarog}}\email{btwarog@ku.edu}

\author[5]{\fnm{Aaron} \sur{Steinhauer}}\email{steinhau@geneseo.edu}

\author[6]{\fnm{Jeremy R.} \sur{King}}\email{jking2@clemson.edu}

\affil*[1]{\orgdiv{Tsung-Dao Lee Institute}, \orgname{Shanghai Jiao Tong University}, \orgaddress{\city{Shanghai}, \postcode{200240}, \country{China}}}

\affil*[2]{\orgdiv{Department of Astronomy}, \orgname{Tsinghua University}, \orgaddress{\city{Beijing}, \postcode{100084}, \country{China}}}

\affil[3]{\orgdiv{Department of Astronomy}, \orgname{Indiana University}, \orgaddress{\street{727 East 3rd Street}, \city{Bloomington}, \postcode{47408}, \state{IN}, \country{USA}}}

\affil[4]{\orgdiv{Department of Physics and Astronomy}, \orgname{University of Kansas}, \orgaddress{\street{1251 Wescoe Hall Dr.}, \city{Lawrence}, \postcode{66045}, \state{KS}, \country{USA}}}

\affil[5]{\orgdiv{Department of Physics and Astronomy}, \orgname{State University of New York}, \orgaddress{\city{Geneso}, \postcode{14454}, \state{NY}, \country{USA}}}

\affil[6]{\orgdiv{Department of Physics and Astronomy}, \orgname{Clemson University}, \orgaddress{\street{118 Kinard Laboratory}, \city{Clemson}, \postcode{29634-0978}, \state{SC}, \country{USA}}}

\abstract{The ``Li-Dip” is an unexpected, striking, and highly non-standard anomaly of severe lithium depletion observed in mid-F dwarf stars, which has puzzled astronomers for nearly 40 years.  Mechanisms proposed to explain the Li-Dip include effects related to rotation, magnetic fields, diffusion, gravity waves, and mass loss. The critical question became, which, if any, might be realistic? Here we show that mixing due to shear induced by stellar angular momentum loss is the unique mechanism driving the Li depletion. Each mechanism leaves a different signature in the subsurface Li distribution. The deepening surface convection zones of subgiants of NGC 188 evolving out of the Li-Dip dredge up the sub-surface material and thus reveal the signature of the responsible mechanism, rotation. Beryllium and boron data have also favored rotational mixing; however, these elements can be extremely difficult or impossible to observe. Our highly complementary approach provides fresh and very feasible perspectives on using Li to probe poorly understood physical mechanisms acting below the stellar surface, thereby improving fundamental understanding of stellar evolution. Rotational mixing may be the dominant mechanism that depletes Li in a wide range of Solar-type stars, including in the Sun. Possible connections to Big Bang Nucleosynthesis are discussed.}

\maketitle

Inside main sequence stars (dwarfs), Lithium (Li), Beryllium (Be), and Boron (B) survive only in the outermost, coolest layers, to progressively greater depths. Their observed surface abundances thus provide invaluable information about physical mechanisms occurring below.  The ``standard” theory of stellar evolution (SSET; \citealp{1967ARA&A...5..571I}, \citealp{1990ApJS...73...21D} \citealp{2017AJ....153..128C}), ignores complicated but potentially interesting effects due to rotation, magnetic fields, diffusion, and mass loss. The SSET predicts that G dwarfs like the Sun and even more so K dwarfs will have depleted some Li due to convective mixing to deep, hot layers where Li is destroyed by energetic protons — and will have done so only during the pre-main sequence. More massive stars like F and A dwarfs will have suffered a negligible Li depletion. In sharp contradistinction to the SSET, a severe Li depletion was discovered in mid-F dwarfs (\citealp{1986ApJ...302L..49B}, the ``Li-Dip”, Figure 1).  This provided the first astonishingly clear evidence that the SSET was missing important physics. Attempts to identify the missing physics quickly proliferated, and included mechanisms such as diffusion \cite{1986ApJ...302..650M, 1993ApJ...416..312R}, mass loss \cite{1990ApJ...359L..55S}, and slow mixing due to effects either of gravity waves \cite{1991ApJ...377..268G} or rotation \cite{1989ApJ...338..424P, 1990ApJS...74..501P}. During the next two decades it became clear that G and K dwarfs also deplete Li during the main sequence \cite{1999MNRAS.304..821J, 2002MNRAS.336.1109J}, which is highly non-standard, and more recently it was discovered that A dwarfs also do so \cite{2019AJ....158..163D}. Although the standard model of the current Sun, a G2V dwarf, agrees very well with helioseismology (e.g., \citealp{1995RvMP...67..781B, 2004PhRvL..92l1301B}), the predicted Li depletion of only a factor of three (``X” in Figure 1) underestimates the actual depletion by a factor of over 50! This huge discrepancy points to an enormous failure of the SSET, and we need to understand the secret life that the Sun has led. So, nearly {\it all} dwarfs for which it is possible to observe the surface Li abundance deplete Li over time in a non-standard way. Understanding the physical cause of these Li depletions is thus of fundamental importance to improving our knowledge of stellar evolution. G and K dwarfs and, very surprisingly, A dwarfs, also spin down during the main sequence, which suggests a possible connection between stellar spindown and Li depletion \cite{2019AJ....158..163D}. For the Li-Dip, Be and B data discriminate between scenarios because whereas all the proposed mechanisms create a Li-Dip, the ratios between Li, Be, and B differ from mechanism to mechanism. The correlations between Li and Be depletion in dwarfs \cite{1998ApJ...498L.147D, 2004ApJ...613.1202B} and between Be and B depletion \cite{1998ApJ...492..727B, 2005ApJ...621..991B, 2016ApJ...830...49B} have argued strongly against diffusion, mass loss, and slow mixing due to gravity waves, and favor a specific kind of rotationally induced mixing, namely, that arising in models where angular momentum loss, which results in stellar spindown, triggers an internal shear instability that results in internal angular momentum transport and mixing (``Yale models”, \citealp{1989ApJ...338..424P, 1997ApJ...488..836D, 2015MNRAS.449.4131S}).  In this paper we show it is possible to use Li, alone, in a very different and highly complementary way to discriminate between scenarios, by studying Li in subgiants evolving out of the Li-Dip.

\begin{figure}
	\renewcommand\thefigure{Figure 1}
	\centering
	\includegraphics[width=1.0\textwidth]{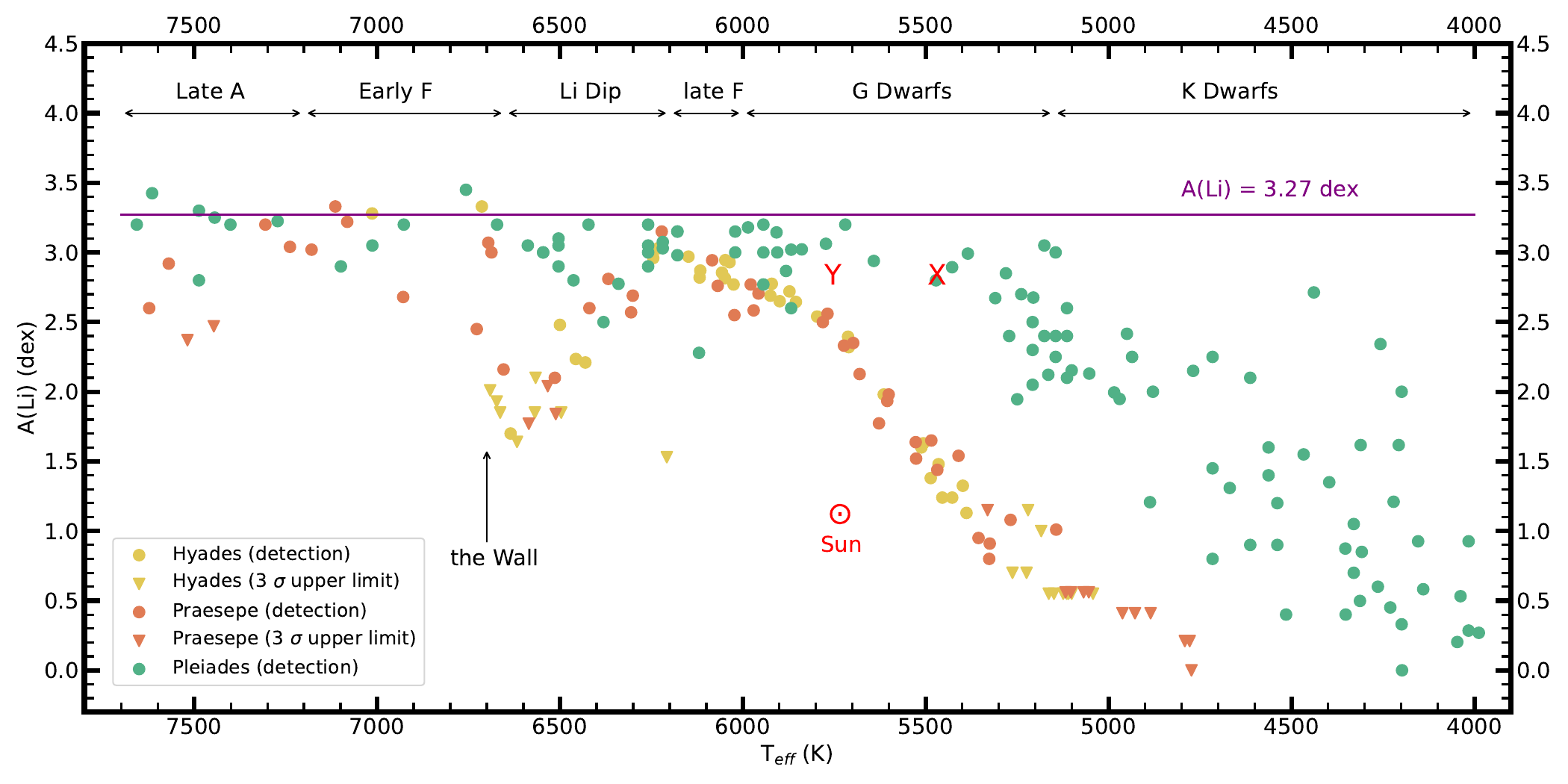}
	\caption{A star cluster contains a coeval set of stars of determinable age that span a variety of stellar masses. Each cluster in this figure thus provides a snapshot of how, at a given age, A(Li) depends on mass, or equivalently $T_{\rm eff}$ for main sequence stars. Only single members are shown, as binaries can lead to various errors and misinterpretations \cite{1993ApJ...415..150T, 2023ApJ...952...71S}. The SSET predictions are approximately consistent with the lower envelope in the Pleiades (age $\sim$ 120 Myr). However, the SSET fails to explain the observed scatter, especially in cooler Pleiads \cite{2017AJ....153..128C, 2018MNRAS.476.3245J}, and strikingly fails to explain the main sequence Li depletion exemplified by Hyades and Praesepe (ages $\sim$ 650 Myr).  Also shown are the meteoritic abundance at A(Li) = 3.27 dex \cite{2021SSRv..217...44L}, which is the presumed initial Solar abundance, the value predicted from the SSET for the current Sun (age $\sim$ 4.5 Gyr, \citealp{1997ARA&A..35..557P}, ``Y”; ``X” indicates this A(Li) at the $T_{\rm eff}$ the Sun would have had at 120 Myr using Yonsei-Yale ($Y^2$) isochrones \cite{2001ApJS..136..417Y}, to compare to the Pleiades), and the actual current Solar A(Li) \cite{1997AJ....113.1871K}, which lies fully a factor of 150 below the meteoritic abundance. Pleiades data are from \cite{2007PhDT.........2M} and \cite{2018AA...613A..63B}, as placed on consistent $T_{\rm eff}$ and A(Li) scales by \cite{2023ApJ...952...71S}, while Hyades and Praesepe data are from \cite{2017AJ....153..128C} and include their own data plus those compiled therein from \cite{1993ApJ...415..150T, 1988ApJ...332..410B, 1986ApJ...302L..49B, 1990AJ.....99..595S, 1989A&A...220..197B}, which were placed by them on the same consistent $T_{\rm eff}$ and A(Li) scales.}
\end{figure}

The Li preservation region is a key discriminator because each of the proposed mechanisms creates a different Li profile (dependence of the Li abundance with depth, \citealp{2000ApJ...544..944S}).  It is instructive to use the SSET as a reference: In the SSET, nearly the entire Li preservation region is radiative, so no mixing occurs except inside the surface convection zone (SCZ). The SCZ is much shallower than the Li preservation region so convection plays almost no role in modifying the Li profile. The Li abundance remains constant as a function of depth until the temperature is high enough to destroy Li; then the Li abundance drops very sharply with depth (the ``Li preservation boundary”). In the mass loss scenario, nearly all of the Li preservation region is lost, so that the Li abundance already drops very sharply with depth immediately below the very shallow SCZ. In the diffusion scenario, the downward diffusion time scales increase with depth, which causes Li to increasingly pile up with depth until the Li preservation boundary is reached. In the Yale models, slow rotational mixing reduces the amount of Li in the Li preservation region compared to the SSET, and, importantly, since the efficiency of the shear-induced mixing decreases with depth, the Li profile is steeper than in the SSET but shallower than in the mass loss scenario.

Subgiants evolving out of the Li-Dip can reveal which of these strikingly different Li profiles is most realistic \cite{2000ApJ...544..944S}, if any!  This is because as subgiants evolve to lower surface temperatures ($T_{\rm eff}$), their SCZs deepen considerably, eventually dredging up the entire Li preservation region. Thus, the Li-$T_{\rm eff}$ relation in star clusters with well-populated subgiant branches determines the true shape of the Li profile. Note that although subgiants evolve to a large range of $T_{\rm eff}$, the entire subgiant branch in a given cluster has evolved out of a very small range of $T_{\rm eff}$ (mass) of the main sequence. In the SSET, the surface A(Li) stays constant from the main sequence turnoff to lower $T_{\rm eff}$ until the SCZ reaches the Li preservation boundary. As the SCZ deepens further with lower $T_{\rm eff}$ and mixes the now fixed amount of Li with regions containing no Li, A(Li) decreases through ``subgiant dilution” until the SCZ contains most of the mass of the star, on the red giant branch. The total amount of dilution is nearly 2 dex \cite{2010A&A...522A..10C}. In the mass loss scenario, A(Li) declines dramatically with lower $T_{\rm eff}$ starting immediately with evolution past the turnoff. In the diffusion scenario, A(Li) initially increases with lower $T_{\rm eff}$ as the SCZ dredges up increasingly piled up Li, until subgiant dilution. For the Yale rotational scenario, the Li-$T_{\rm eff}$ trend has some negative slope but not as steep as in the mass loss scenario.

Li data from star cluster M67 were possibly consistent with Yale rotational mixing, and provided strong evidence against diffusion, but could not rule out mass loss if, for example, the model stellar $T_{\rm eff}$  scale was too high by of order 200K \cite{2000ApJ...544..944S}. This ambiguity was removed when Be data in M67 were combined with the Li data, ruling out mass loss and strongly supporting rotational mixing \cite{2020ApJ...888...28B}. However, Be data can be extremely challenging to obtain, requiring much observing time on only a few of the world’s largest telescopes that are suitably equipped with a UV-capable high resolution spectrograph. For example, the M67 Be data required data acquisition at the Keck I 10-meter telescope over a period of 4 years, for only 9 stars; this is the most efficient telescope in the world for stellar Be work. B data are even more challenging to obtain, requiring use of the Hubble Space Telescope. Moreover, Be observations are limited to far brighter stars than can be observed effectively for Li, and B observations are limited to even brighter stars. For example, M67 subgiants are much too faint for B observations. It is therefore of great interest to develop methods where Li data alone, without additional information from Be and/or B data, suffice to give us a reasonably full picture. The Li observations of subgiants on NGC 188 do just that, while also opening up great opportunities for the future.

NGC 188 is slightly older (age $\sim$ 6 Gyr, \citealp{2022MNRAS.513.5387S}) than M67 (age $\sim$ 4 Gyr, \citealp{2020AJ....159..246S}), so its subgiants evolve out of a slightly cooler portion of the Li-Dip, and are slightly less massive than those in M67. To create a Li-Dip, the Li profile in the mass loss scenario remains very steep as a function of depth, and in the diffusion scenario the Li abundances again increase at first with depth until the Li preservation boundary is reached. For rotational mixing, the Li profile is shallower than in M67 because of the smaller stellar mass \cite{1997ApJ...488..836D}. The implications for subgiants in NGC 188 evolving to lower $T_{\rm eff}$ are that, a) for mass loss, the Li abundances will again decline very steeply, b) for diffusion, the Li abundances will again increase until dilution begins, and c) for rotational mixing, the Li-$T_{\rm eff}$ relation will be shallower than that in M67.

Figure 2 shows Li abundances for subgiants from the turnoff to undetectably low levels for M67 \cite{2000ApJ...544..944S} and for NGC 188 (presented in this study). The Li-$T_{\rm eff}$ relation is substantially shallower for NGC 188 than for M67.  Such a shallow slope is inconsistent with mass loss.  The slope is also decreasing with lower $T_{\rm eff}$ instead of increasing, inconsistent with diffusion. But such a slope is consistent with rotational mixing, especially the fact that the slope is shallower than in M67. Thus, the Li-$T_{\rm eff}$ relation for NGC 188 subgiants supports rotational mixing and argues against mass loss and diffusion.

Originally it was Li data alone that signaled the striking effects of highly non-standard physical mechanisms acting in the outer layers of stars during the main sequence to deplete their surface Li abundance, creating, for example, the Li-Dip.  Since then, the strongest evidence identifying rotationally induced mixing as the dominant responsible mechanism instead of other proposed ones such as diffusion or mass loss has come from the correlated depletion of Li with depletion Be and B, and with stellar spindown.  Finally, and importantly, nearly 40 years after the discovery of the Li-Dip, we have come full circle:  these Li data alone point to rotationally induced mixing as the unique, dominant Li depletion mechanism creating the Li-Dip.

The $T_{\rm eff}$ range (6200 – 6000 K) of the NGC 188 subgiant progenitors coincides, in part, with the Li–$T_{\rm eff}$ Plateau observed in metal-poor dwarfs \cite{1982Natur.297..483S, 1982A&A...115..357S}. Rotational mixing depleted the Li of these progenitors by about a factor of 10, as suggested by current turnoff stars (Figures 2, 3) and a presumed initial A(Li) near the meteoritic A(Li), or slightly higher to account for Galactic Li production \cite{2011PhDT.......192C, 2020A&A...640L...1R} and NGC 188’s super-solar metallicity. It is then plausible that metal-poor dwarfs are depleted by only about a factor of 3, due to their lower mass and potentially lower initial angular momentum. Such a depletion could reconcile the observed level of the Li-$T_{\rm eff}$ plateau with the A(Li) predicted in standard models of Big Bang Nucleosynthesis \cite{1996RPPh...59.1493S, 2020JCAP...03..010F}. 

\begin{figure}
	\renewcommand\thefigure{Figure 2}
	\centering
	\includegraphics[width=0.7\textwidth]{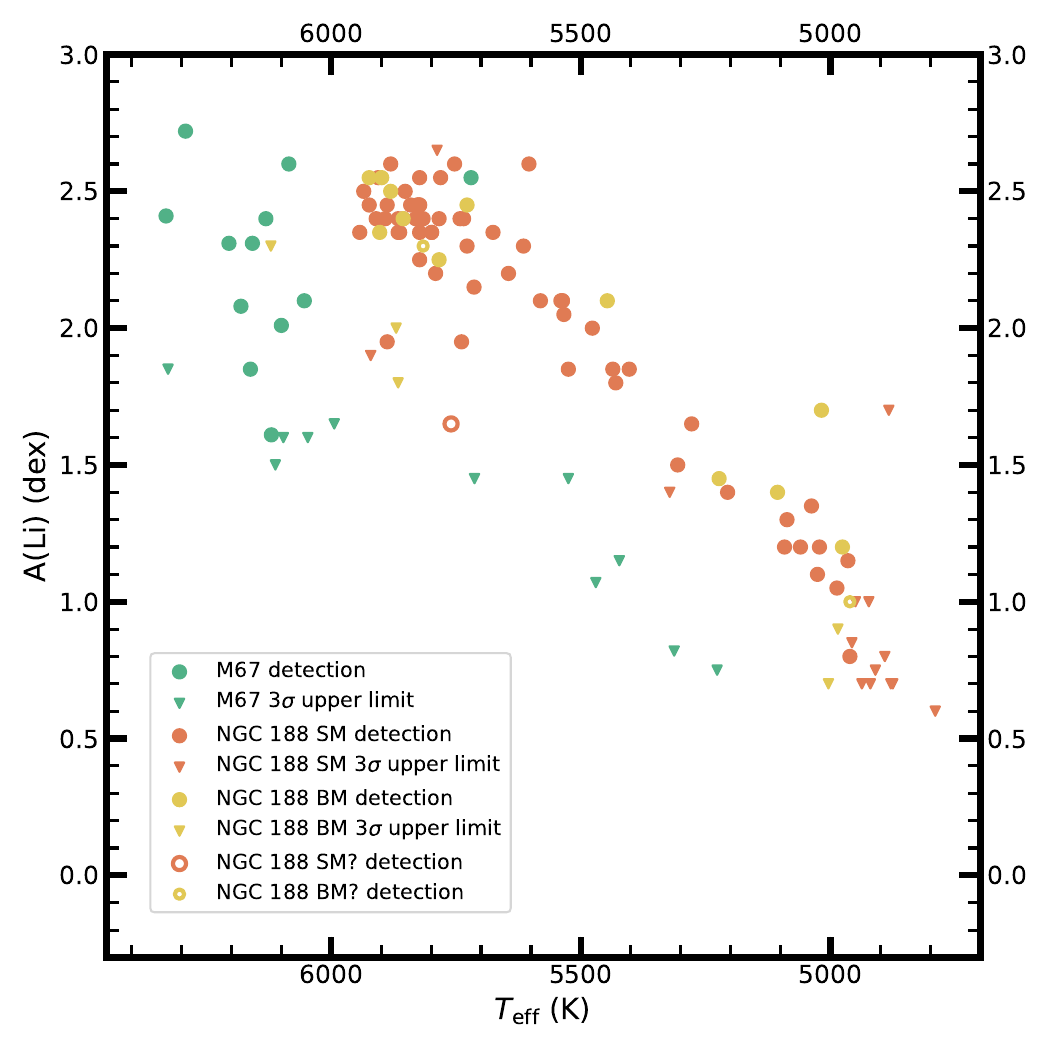}
	\caption{A(Li) versus $T_{\rm eff}$ for turnoff stars and subgiants in M67 and NGC 188. The single members (SM), binary members (BM), single likely members (SM?), and binary likely members (BM?) of NGC 188 are shown. Members of M67 from \cite{2000ApJ...544..944S}, with membership probability $>$ 0.5 from \cite{2018A&A...618A..93C}, are also shown.}
\end{figure}

\section*{Methods}

\bmhead{Observations}

Data for candidate members of NGC 188 ranging from dwarfs with less than 1 solar mass up through the red giant branch were obtained using the multi-fiber Hydra spectrograph mounted on the WIYN 3.5-m telescope, by using multiple configurations during different observing runs in 1995 November, 1996 April, 1997 June, 2001 February and March, 2002 April, and 2017 December.  The echelle 316 l/mm grating was used, which provided a resolution of approximately R $\sim$ 13,000 with the blue fibers and R $\sim$ 17,000 with the red fibers, covering a range of approximately 400 \AA\ centered near 6650 \AA. In 2001 March and 2002 April, we utilized the 31.6 l/mm KPNO coude grating, covering a wavelength range of 6700-6730 \AA, to take advantage of its higher throughput efficiency and slightly higher resolution (R $\sim$ 19,000).  This grating works well for studying the Li I feature at 6707.78 \AA, but has an insufficient number of Fe I lines for determination of metallicity; therefore, metallicity has been determined using data from the 316 l/mm grating \cite{2022MNRAS.513.5387S}. 

\bmhead{Sample Selection}

To focus on the topic of this study, we have carefully selected 96 turnoff and subgiant members from NGC 188, including only turnoff stars more evolved than $V = $15.4 mag, and subgiants as bright as $V =$ 14.16 mag. Assuming a distance modulus (m-M) = 11.48 mag and E(B-V) = 0.09 mag \cite{2022MNRAS.513.5387S}, and applying the extinction relation $A_V$ = 3.1 E(B-V) \cite{1989ApJ...345..245C}, we transform the apparent $V$ to $M_V$, and show the selected stars in Extended Figure 1. For comparison, we also show members of M67 \cite{2000ApJ...544..944S}, transforming them assuming (m-M) = 9.75 mag and E(B-V) = 0.04 mag \cite{2020AJ....159..246S}. A study of subgiants and giants brighter than $V =$ 14.16 mag and the apparent production of Li in some of these stars was presented in \cite{2022MNRAS.513.5387S}, and main sequence stars fainter than $V = $15.4 mag will be discussed in future work.

\bmhead{Data Reductions and Membership/Multiplicity}

Following similar procedures as described in \cite{2022MNRAS.513.5387S}, we removed the instrumental signature from the spectra, calculated radial velocities for each night, and used the combined spectra along with Gaia DR2 \cite{2018AA...616A...1G} proper motion and parallax data, to assess multiplicity and membership of each individual star. Using similar precepts, we designated  76 stars as SM (``single member"), 17 as BM (``binary member"), 1 as SM? (``single likely member"), and 2 as BM? (``binary likely member"). The fundamental stellar parameters are included in supplementary Table 1, and the radial velocities for individual nights are included in supplementary Table 2.

\bmhead{Stellar Atmospheres}

The procedures follow closely those described in \cite{2022MNRAS.513.5387S}. We have adopted their [Fe/H] of +0.064 $\pm$ 0.018 dex, and utilized their choice of a 6.3 Gyr $Y^2$ \cite{2004ApJS..155..667D} isochrone with E(B-V) = 0.09 mag. However, given that our sample comprises both turnoff stars and subgiants, we tailored our approach by applying their color -- $T_{\rm eff}$ and microturbulence relations for subgiants and giants with $T_{\rm eff} < 5535$ K ($(B-V)_0 > 0.706$ mag), and those of \cite{2023ApJ...952...71S} for subgiants and dwarfs with $T_{\rm eff} > 5535$ K ($(B-V)_0 < 0.706$ mag).

\bmhead{Li abundances}

The procedures closely follow those described in \cite{2022MNRAS.513.5387S} and \cite{2023ApJ...952...71S}. Briefly, for each of the 96 stars we create synthetic spectra to derive lithium abundances, A(Li), and to distinguish between Li detections from $3\sigma$ upper limits guided by the precepts in \cite{1993ApJ...414..740D}. The spectroscopic abundance scale, A(X), is defined as 12 + log(N$_{Li}$/N$_H$), where N$_X$ is number of nuclei of species X and N$_H$ is the number of hydrogen nuclei. The final A(Li) detections and upper limits are shown in Figure 2 and included in Extended Table 1.

\backmatter

\bmhead{Data Availability}

The spectra acquired from the WIYN telescope can be shared upon request to the corresponding author. All other data supporting this article can be found within the article itself and its online supplementary material.

\bmhead{Code Availability}

The MOOG stellar line analysis program can be accessed from the website: \url{https://www.as.utexas.edu/~chris/moog.html}. Stellar model atmospheres are obtained from the following source: \url{http://kurucz.harvard.edu/grids.html}.

\bmhead{Acknowledgments}

Q.S. thanks support from the Shuimu Tsinghua Scholar Program. NSF support for this project was provided to C.P.D. through
grant AST-1909456. We also thank the WIYN 3.5 m staff for helping us obtain excellent spectra.

This work has made use of data from the European Space Agency (ESA) mission Gaia (\url{https://www.cosmos.esa.int/gaia}), processed by the Gaia Data Processing and Analysis Consortium (DPAC, \url{https://www.cosmos.esa.int/web/gaia/dpac/consortium}). Funding for the DPAC has been provided by national institutions, in particular, the institutions participating in the Gaia Multilateral Agreement. 

\bmhead{Author contributions}

Q.S. and C.P.D. designed the paper, conducted observations, analyzed the spectra, and prepared the paper. B.A.T. and B.J.A.T. helped with the scientific interpretation of the data. A.S and J.R.K also conducted observations. All authors thoroughly reviewed, provided feedback on, and reached a consensus on the manuscript.

\bmhead{Competing interests}

The authors have no competing interests to declare.

\bmhead{Supplementary information}

\begin{table*}
	\renewcommand\thetable{Extended Table 1}
	\caption{M67 A(Li) data}
	\begin{tabular}{ccccccc}
		\hline
		ID$^1$ & RA$^1$ (J2000) & DEC$^1$ (J2000) & $V^1$ & $B-V^1$ & $T_{\rm eff}^2$ & A(Li)$^2$ \\
		 & $^h\ ^m\ ^s$ & $^{\circ}$ ' '' & mag & mag & K & dex  \\
		\hline
		5123 & 132.67453 & 11.61483	& 12.647 & 0.685 & 5713	& $<$1.37 \\
		5041 & 132.64307 & 11.66580	& 12.629 & 0.594 & 6047	& $<$1.50 \\
		5042 & 132.64119 & 11.77392	& 12.856 & 0.531 & 6292	&   2.62 \\
		5118 & 132.66988 & 11.79610	& 12.818 & 0.521 & 6331	&    2.31 \\
		5061 & 132.65271 & 11.81544	& 13.025 & 0.574 & 6123	&    2.26 \\
		5191 & 132.69851 & 11.74733	& 12.700 & 0.482 & 6490	& $<$2.22 \\
		5228 & 132.70811 & 11.82027	& 12.931 & 0.851 & 5227	& $<$0.77 \\
		5362 & 132.75447 & 11.83643	& 12.725 & 0.739 & 5525	& $<$1.37 \\
		5284 & 132.72659 & 11.94134	& 12.844 & 0.522 & 6327 & $<$1.75 \\
		5219 & 132.70275 & 12.00243	& 12.797 & 0.553 & 6205 &    2.21 \\
		5675 & 132.83399 & 11.77831	& 12.755 & 0.559 & 6181 &    1.98 \\
		5562 & 132.81163 & 11.79006	& 12.805 & 0.572 & 6131 &    2.30 \\
		5644 & 132.82737 & 11.82270	& 12.647 & 0.608 & 5994 & $<$1.55 \\
		5573 & 132.81406 & 11.83739	& 12.823 & 0.565 & 6158	&    2.21 \\
		5825 & 132.86439 & 11.89076	& 12.756 & 0.592 & 6054 &    2.00 \\
		6076 & 132.92490 & 11.72707	& 12.69	 & 0.564 & 6162 &    1.75 \\
		6107 & 132.93353 & 11.77351	& 12.75	 & 0.758 & 5470 & $<$1.01 \\
		5993 & 132.90023 & 11.77605	& 12.722 & 0.683 & 5720 &    2.47 \\
		6177 & 132.95829 & 11.82540	& 12.647 & 0.581 & 6096 & $<$1.49 \\
		6077 & 132.92196 & 11.90817	& 12.76	 & 0.580 & 6100 &    1.90 \\
		5996 & 132.89769 & 11.96578	& 12.826 & 0.775 & 5423 & $<$1.10 \\
		6408 & 133.04736 & 11.76044	& 12.889 & 0.817 & 5313 & $<$0.81 \\
		6224 & 132.97245 & 11.80585	& 12.705 & 0.575 & 6120 &    1.51 \\
		6313 & 132.99856 & 11.88273	& 12.79	 & 0.584 & 6085 &    2.49 \\
		5951 & 132.88859 & 11.81431	& 12.767 & 0.577 & 6112 & $<$1.39 \\
		\hline
	\multicolumn{7}{p{.9\textwidth}}{Notes 1: ID, $V$, and $B-V$ values are sourced from \cite{1990BAAS...22.1288M}. The original RA and DEC, initially in B1950, have been converted to J2000 using the astropy package \cite{2022ApJ...935..167A}.} \\
	\multicolumn{7}{p{.9\textwidth}}{Notes 2: The original $T_{\rm eff}$ and A(Li) values have been recalibrated to our $T_{\rm eff}$ scale, enabling a direct comparison with our dataset. Only stars with M67 cluster membership $>$ 0.5 from \text{\cite{2018A&A...618A..93C}} are shown.} \\
	\end{tabular}
	\label{tab:OC}
\end{table*}

\begin{figure}
	\renewcommand\thefigure{Extended Figure 1}
	\centering
	\includegraphics[width=1.0\textwidth]{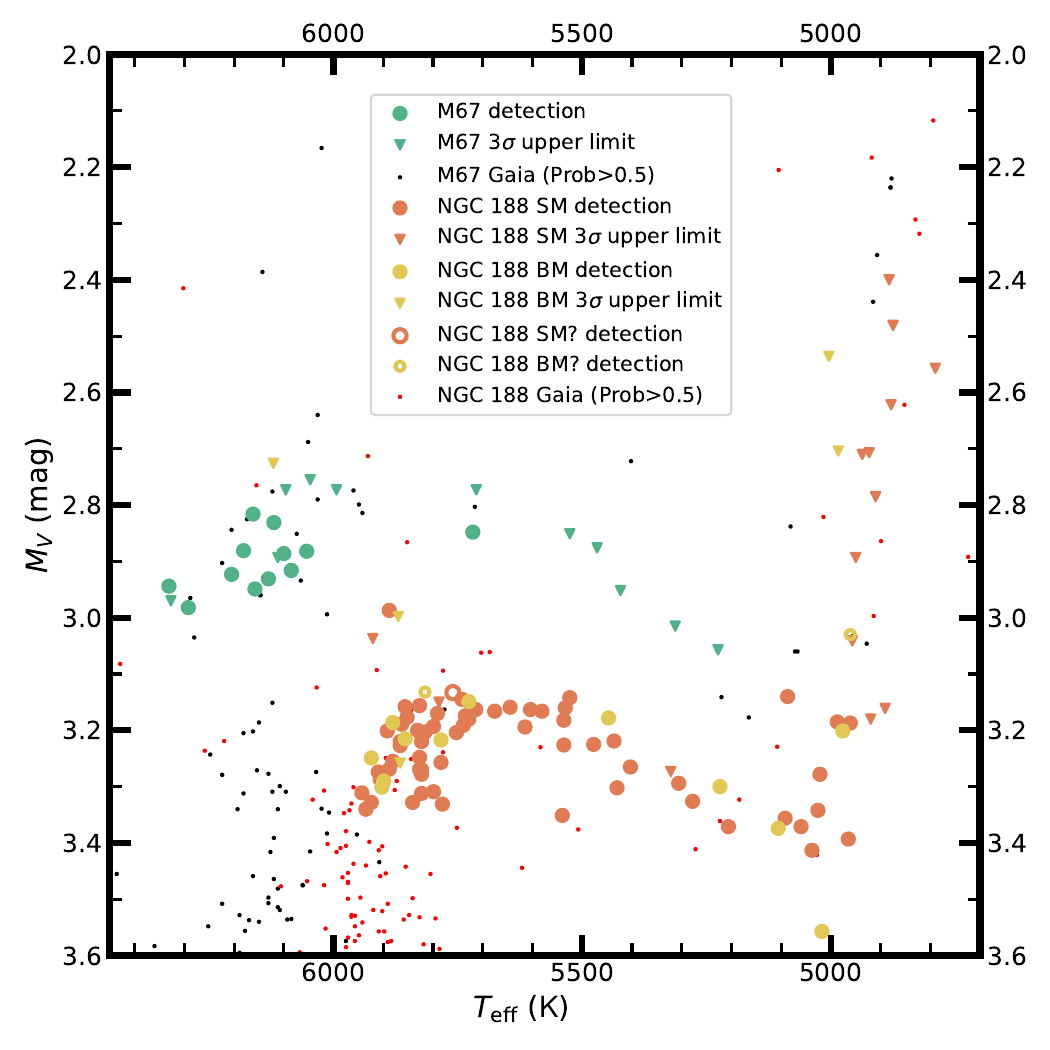}
	\caption{Hertzsprung-Russell diagram of stars from M67 and NGC 188. Small dots indicate members of M67 and NGC 188, selected based on a membership probability greater than 0.5 from \cite{2018A&A...618A..93C}. The $V$ and $B-V$ values for Gaia members in M67 are retrieved by cross-matching with data from \cite{1990BAAS...22.1288M}, and for NGC 188 are by cross-matching with data from \cite{2003AJ....126.2922P}. Larger symbols indicate stars selected in this study.  Disks show stars with Li detections whereas inverted triangles show stars with Li upper limits.  Single members (SM) and single likely members (SM?) of NGC 188 are in orange whereas binary members (BM) and binary likely members (BM?) are in yellow. Members of M67 from \cite{2000ApJ...544..944S}, with membership probability $>$ 0.5 from \cite{2018A&A...618A..93C}, are shown in green.}
\end{figure}

\noindent \textbf{Correspondence and requests for materials} should be directed to Q. Sun.

\bibliography{sn-bibliography}

\end{document}